\documentclass[numberedappendix,onecolumn]{emulateapj}
\pdfoutput=1

\usepackage{amsmath}

\usepackage{graphicx}
\usepackage{natbib}
\usepackage{color}
\usepackage[bookmarks=true]{hyperref}

\newcommand{\D}[2]{\ensuremath{\frac{d{#1}}{d{#2}}}}
\newcommand{\Ht}{\ensuremath{\mathcal{H}}}
\newcommand{\Lt}{\ensuremath{\mathcal{L}}}
\newcommand{\eqnref}[1]{(\ref{#1})}

\defcitealias{NB07}{NB07}

\begin{document}

\title{Galaxy Modeling with Compound Elliptical Shapelets}
\author{James Bosch}
\affil{Physics Department, University of California, Davis}
\email{jbosch@physics.ucdavis.edu}

\begin{abstract}
  Gauss--Hermite and Gauss--Laguerre (``shapelet'') decompositions of
  images have become important tools in galaxy modeling, particularly
  for the purpose of extracting ellipticity and morphological information
  from astronomical data.  However, the standard shapelet basis functions
  cannot compactly represent galaxies with high ellipticity or large
  S\'{e}rsic index, and the resulting underfitting bias has been
  shown to present a serious challenge for weak-lensing methods based on
  shapelets.  We present here a new convolution relation and a compound
  ``multiscale'' shapelet basis to address these problems, and
  provide a proof-of-concept demonstration using a small sample of
  nearby galaxies.
\end{abstract}

\section{Introduction}
\label{sec:introduction}

With the advent of large photometric surveys, efficient, automated
techniques for galaxy modeling have become an important part of
astronomical data analysis.  Models based on a linear combination of
Gauss-Hermite or Gauss-Laguerre functions (also known as Cartesian or
polar ``shapelets'', respectively) have become
particularly important in weak-lensing ellipticity measurement, and
have also been used for morphological classification and in building
realistic mock image data (see, for instance,
\citet{BJ02,R03,MR05,Massey2004,KellyMcKay2004,Andrae2010}) .

Because the standard shapelet bases consist of polynomials weighted
by a circular Gaussian, they are a good first-order approximation to typical
galaxy and point-spread function (PSF) morphologies.  More
importantly, a shapelet expansion can be convolved analytically, which
substitutes a simple linear algebra operation for what can otherwise
be the most computationally expensive aspect of a modeling algorithm.

Recent work has highlighted the drawbacks of standard shapelet-based galaxy
modeling, however, and demonstrated that even high-order shapelet
expansions are often poor representations of real galaxies.
\citet{Melchior2010} have shown that these deficiencies can introduce
serious systematic ellipticity biases in shapelet-based weak lensing
measurements.  In particular, shapelets cannot reproduce the sharp
core and broad wings of galaxies with high S\'{e}rsic indices, and become
increasingly distorted at high ellipticities.

We propose here a shapelet-based modeling technique that
can much more compactly represent real galaxies, while preserving the
lossless analytic convolution and other useful properties of the
standard shapelet expansion.  By combining multiple low-order shapelet
expansions with different scales, our technique can simultaneously
represent the extended wings and cuspy cores of real galaxies.  We
also present a new convolution relation for the ellipse-transformed
shapelet expansion of \citet{NB07} (hereafter \citetalias{NB07}),
enabling lossless high-ellipticity shapelet modeling and eliminating
one source of multiplicative shear bias present in shapelet-based weak
lensing methods.

In Section~\ref{sec:limitations}, we summarize the limitations of
standard shapelet modeling techniques.  In
Section~\ref{sec:convolution}, we derive an exact convolution relation
for elliptical shapelets, and we discuss the combination of multiple
shapelet expansions into a single compound expansion in
Section~\ref{sec:modeling}.  We provide a simple demonstration of
the compound shapelet technique using the \citet{Frei1996} sample of
nearby galaxies in Section~\ref{sec:demos}, and conclude in
Section~\ref{sec:conclusion}.

\section{Limitations of standard shapelet techniques}
\label{sec:limitations}

\subsection{Ellipticity}
\label{sec:limitations:ellipticity}

The standard Cartesian shapelet basis functions are formed from the
product of two one-dimensional Gauss--Hermite functions:
\begin{align}
  \Phi_{\mathbf{n}}(\mathbf{x}) &= \phi_{n_1}(x_1) \phi_{n_2}(x_2)\\ &=
  \frac{H_{n_1}(x_1) H_{n_2}(x_2) e^{-\frac{1}{2}(x_1^2 + x_2^2)}} 
       {\sqrt{2^{n_1+n_2}\,\pi\,n_1!\,n_2!}} .
  \label{eqn:2d-basis}
\end{align}
This is a clearly an expansion about a circular Gaussian, and is
naturally poor at representing functions with high ellipticity.
However, we can create an elliptical expansion by transforming the
coordinate grid by the inverse of the transform that maps the unit
circle to the desired ``basis ellipse'' \citepalias{NB07}: 
\begin{equation}
  \Phi_{\mathbf{n}}\left(\mathbf{x}|\mathbf{e}\right) 
  = \Phi_{\mathbf{n}}(\mathbf{S}_e^{-1}\mathbf{x})
  \label{eqn:elliptical-shapelets}
\end{equation}
\begin{equation}
  \mathbf{S}_e = \left[\begin{array}{c c}
    \cos\theta & -\sin\theta \\
    \sin\theta & \cos\theta
    \end{array}\right]
  \left[\begin{array}{c c}
      a & 0 \\
      0 & b
    \end{array}\right].
\end{equation}
where $a$, $b$, and $\theta$ are the semi-major axis, semi-minor axis,
and position angle of the ellipse, respectively.\footnote{An arbitrary
  rotation can also be applied on the right side, as this corresponds
  to a rotation of the unit circle; \citetalias{NB07} rotate by
  $-\theta$ to make $\mathbf{S}$ symmetric,
  but we prefer to omit the rotation in order to align the
  shapelet expansion with the axes of the ellipse rather than the
  original coordinate grid.}
This transformation is exact, but the standard shapelet convolution
formula \citep[see][]{R03} applies only to shapelet expansions with
identical ellipticities; this limits analytic convolution to the
case where both the galaxy and the PSF are approximately circular.

\begin{figure*}
  \begin{center}
    \includegraphics[width=\textwidth]{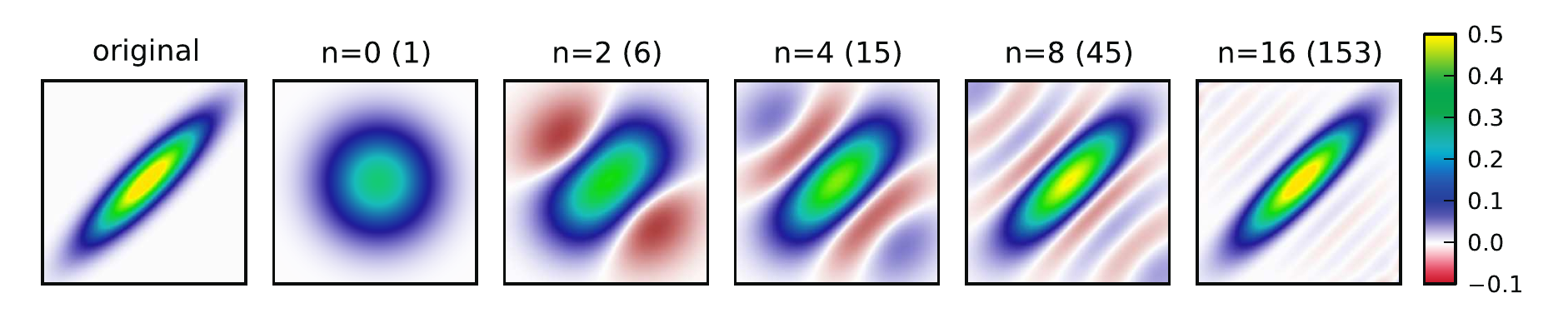}
  \end{center}
  \caption{Circular shapelet approximations of increasing order $n$ to
    an elliptical Gaussian with an axis ratio of $1/4$.  The number of
    basis functions ($\frac{1}{2}(n+1)(n+2)$) used in each approximation
    is given in parenthesis.  Because the number of basis functions
    increases so dramatically at high order, the order of the
    expansion must be kept low for computationally feasible
    techniques, but low-order expansions produce distorted
    representations of even simple high-ellipticity
    morphologies.
    \vspace{2cm}
    \label{fig:ellipticity}}
\end{figure*}

Instead, most shapelet techniques make use of the
shapelet-space shear operator, which can be represented by a matrix
multiplication on the basis vector \citep[see][]{R03,MR05}:
\begin{equation}
  \Phi_{\mathbf{n}}(\mathbf{x}|\mathbf{e}) =
  \sum_{\mathbf{m}}^{\infty}
  \left[\hat{\mathbf{S}}_e\right]_{\mathbf{n},\mathbf{m}}\Phi_\mathbf{m}(\mathbf{x}) .
\end{equation}
This relation must be truncated at finite $\mathbf{m}$ in practice,
however, making the shear operation lossy.  A simple elliptical Gaussian
cannot be represented exactly by a finite circular shapelet expansion,
and the approximation introduces artifacts that become increasingly
severe as the ellipticity increases (Figure~\ref{fig:ellipticity}).  Unless
the average galaxy morphology mimics these distortion patterns (highly
unlikely, given that they involve regions with negative flux), this necessarily
makes the goodness of fit of shapelet models worse, on average, as
ellipticity increases.  While this is most
noticeable at high ellipticities, it is important even
at low ellipticities, as these approximation-induced artifacts
systematically bias shapelet-based lensing techniques \citep{Melchior2010}.  

A better solution is to use elliptical basis functions to model
the galaxy.  To convolve with the PSF, one applies the inverse
shapelet-space shear operator to the PSF to transform it into the
coordinate system in which the galaxy is round \citepalias{NB07}.
However, as the ellipticity of the galaxy model increases, the magnitude of the
inverse shear transform that must be applied to the PSF model also
increases, introducing approximation artifacts in the PSF model.

This tradeoff is advantageous for two reasons.  First, one can often
afford to use a higher-order expansion for the PSF, because the PSF
coefficients are not considered free parameters when fitting an
individual galaxy; the PSF model is determined separately using images
of stars in the same field.  In addition, when the galaxy radius is
larger than the PSF, the undistorted galaxy model plays a
larger role in determining the ellipticity of the convolved model than
the now-distorted PSF model.  Still, this technique does not eliminate
the ``shear artifact'' ellipticity bias; it merely decreases it by
substituting a distorted, approximate PSF model for a distorted,
approximate galaxy model.

\subsection{Radial Profiles}
\label{sec:limitations:profiles}

\begin{figure}
  \begin{center}
  \includegraphics[width=0.5\textwidth]{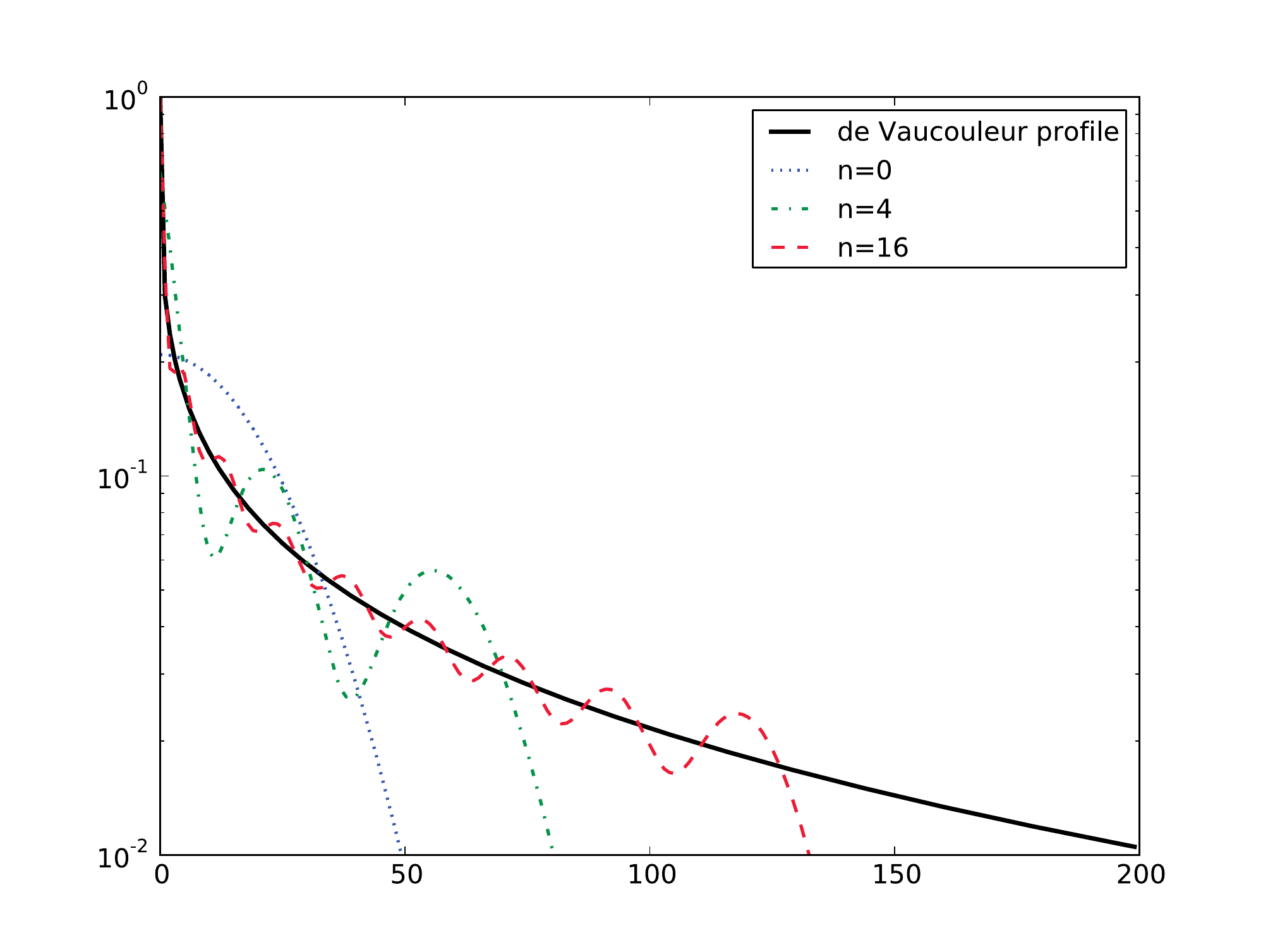}
  \caption{Shapelet approximations to a de Vaucouleur (S\'{e}rsic $\alpha=4$)
    profile.  The radius and flux units are arbitrary, and the fit can
    be tuned to perform well at either small or moderate radii, but not
    both.  Regardless of the scale of shapelet expansion, it will
    always fall off much faster at large radii compared to a pure de
    Vaucouleur profile.  This artificial truncation of the radial
    profile strongly affects flux and size estimates made using the
    model, and can bias the measured ellipticity as
    well, even when the truncation occurs at low surface brightness. 
    \label{fig:profiles}} 
  \end{center}
\end{figure}

Shapelet expansions also have difficulty reproducing realistic galaxy
radial profiles.  The azimuthally averaged radial profile of a galaxy often
follows a S\'{e}rsic law:
\begin{equation}
  \mathrm{flux} \propto e^{-r^{1/\alpha}}
\end{equation}
with $\alpha \ge 1$.  Disk galaxies typically have $\alpha \approx 1$ (an
exponential profile), while spheroidal galaxies often have $\alpha \approx 4$
(the de Vaucouleur profile) or greater.  The shapelet expansion is
based on the Gaussian function ($\alpha=1/2$), and hence has a much softer
core and sharper cutoff at large radii than a S\'{e}rsic profile with 
$\alpha\ge 1$.  In theory, the shapelet basis is complete,
and can absorb these differences in higher-order terms, but in
practice a finite shapelet expansion converges to a S\'{e}rsic model with
high $\alpha$ extremely slowly, producing a clear ``ringing'' pattern in
the approximation.  Figure~\ref{fig:profiles} shows a typical case.

This poses a clear problem for the use of shapelets as a tool for
morphological classification: the slope of the radial profile,
arguably the clearest and most obvious distinction between galaxy
types, is a nonlinear function of high-order terms in shapelet space.
In contrast, it can be easily be estimated with a simple S\'{e}rsic fit
or concentration measurement.

The inability to accurately reproduce realistic radial profiles also
has implications for shear measurement.  Ellipticity estimators based
on model fitting have been shown to produce a multiplicative bias when
the model is a systematically poor fit to the data 
\citep{VB09,Bernstein2010}.  Using mock galaxies with S\'{e}rsic profiles,
\citet{Melchior2010} have shown that this ``underfitting bias'' exists for
shapelets even at relatively high orders.  Even at low-surface
brightness, modeling the wings of galaxies properly can be very
important in shear estimation, as the wings contain
the low-spatial-frequency shape information that is corrupted least by
convolution with the PSF \citep{Bernstein2010}.

A shapelet-like expansion based on general S\'{e}rsic profiles rather
than Gaussians has been proposed as a possible solution to this
problem \citep{Ngan2009}, but these ``S\'{e}rsiclets'' lack many of the
advantages of the shapelet basis, including analytic convolution and
fast numerical evaluation, and thus far have not been used to
construct a practical shear or morphological estimator.

\section{Convolution of elliptical shapelets}
\label{sec:convolution}

\notetoeditor{This section contains many long equations which are
  difficult to break up further into multiple lines.  If possible, it
  should be in a single wide column.}

In this section, we derive an exact analytic convolution formula for
elliptical shapelets.  This provides the ``missing ingredient'' for
making use of the elliptical shapelet basis of Equation
\eqnref{eqn:elliptical-shapelets}
to address the limitations of the standard shapelet basis at high
ellipticity.  The result allows the PSF model and unconvolved galaxy
model to be represented by elliptical shapelets with different basis
ellipses, and determines an optimal basis ellipse for the convolved
shapelet expansion that makes the convolution exact.

Consider a pair of functions and their convolution, each represented
by a finite elliptical shapelet expansion:

\begin{align}
  f(\mathbf{x}) &= \sum_{\mathbf{n}}^{|n|\le N_f} f_{\mathbf{n}}
  \Phi_{\mathbf{n}}\!\left(\mathbf{U}^{-1}\mathbf{x}\right) \\
  g(\mathbf{x}) &= \sum_{\mathbf{n}}^{|n|\le N_g} g_{\mathbf{n}}
  \Phi_{\mathbf{n}}\!\left(\mathbf{V}^{-1}\mathbf{x}\right) \\
  (f * g)(\mathbf{x}) &= \sum_{\mathbf{n}}^{|n|\le N_h} h_{\mathbf{n}}
  \Phi_{\mathbf{n}}\!\left(\mathbf{W}^{-1}\mathbf{x}\right) 
\end{align}

The boldface two-dimensional indices $\mathbf{n}$ each run over pairs
of one-dimensional indices 
$\{n_1,n_2\}$ as in Equation \eqnref{eqn:2d-basis}, but for conciseness we will
define the ``magnitude'' of a vector index as $|n| \equiv n_1 + n_2$, as
this combination will appear often.

The Fourier transforms of the circular
shapelet basis functions are proportional to the basis functions
themselves \citep{R03}, and the Fourier-space ellipse transform is
simply the transpose of the inverse of the real-space ellipse
transform, so

\begin{align}
  \tilde{f}(\mathbf{k}) &= \left|U\right|\!\sum_{\mathbf{n}}^{|n|\le N_f} f_{\mathbf{n}}
  i^{|n|} \Phi_{\mathbf{n}}\!\left(\mathbf{U}^T\mathbf{k}\right) \\
  \tilde{g}(\mathbf{k}) &= \left|V\right|\!\sum_{\mathbf{n}}^{|n|\le N_g} g_{\mathbf{n}}
  i^{|n|}\Phi_{\mathbf{n}}\!\left(\mathbf{V}^T\mathbf{k}\right)\\
  2\pi \tilde{f}(\mathbf{k})\,\tilde{g}(\mathbf{k}) &=
  \left|W\right|\!\sum_{\mathbf{n}}^{|n|\le N_h} h_{\mathbf{n}} 
  i^{|n|}\Phi_{\mathbf{n}}\!\left(\mathbf{W}^T\mathbf{k}\right) \\ 
  &= 2\pi \left|U\right|\left|V\right|\!\sum_{\mathbf{p}}^{|p|\le
    N_f} \sum_{\mathbf{q}}^{|q|\le N_g}
  i^{|p|+|q|} \Phi_{\mathbf{p}}\!\left(\mathbf{U}^T\mathbf{k}\right)
  \Phi_{\mathbf{q}}\!\left(\mathbf{V}^T\mathbf{k}\right)
\end{align}

We can use the orthogonality of the basis functions to isolate the
coefficients of the convolved expansion by multiplying by
$\Phi_{\mathbf{m}}\!\left(\mathbf{W}^T\mathbf{k}\right)$ and
integrating

\begin{align}
  h_{\mathbf{n}} &= 2\pi
  \left|U\right|\left|V\right|\!\sum_{\mathbf{p}}^{|p|\le N_f}
  \sum_{\mathbf{q}}^{|q|\le N_g}
  i^{|p|+|q|-|n|} \int \!d^2\mathbf{k}\,
  \Phi_{\mathbf{p}}\!\left(\mathbf{U}^T\mathbf{k}\right)
  \Phi_{\mathbf{q}}\!\left(\mathbf{V}^T\mathbf{k}\right)
  \Phi_{\mathbf{n}}\!\left(\mathbf{W}^T\mathbf{k}\right)
  \label{eqn:conv-unsimplified}
  \\
  &= 2\pi \left|U\right|\left|V\right|\!\sum_{\mathbf{p}}^{|p|\le N_f}
  \sum_{\mathbf{q}}^{|q|\le N_g}
  i^{|p|+|q|-|n|}\;
  I_{\mathbf{p},\mathbf{q},\mathbf{n}}\!\left(\mathbf{U},\mathbf{V},\mathbf{W}\right)
\end{align}

To compute the integral, it is useful to separate the exponential and
polynomial terms

\begin{align}
  I_{\mathbf{p},\mathbf{q},\mathbf{n}}\!\left(\mathbf{U},\mathbf{V},\mathbf{W}\right)
  &=
  \int \!d^2\mathbf{k}\,
  \Phi_{\mathbf{p}}\!\left(\mathbf{U}^T\mathbf{k}\right)
  \Phi_{\mathbf{q}}\!\left(\mathbf{V}^T\mathbf{k}\right)
  \Phi_{\mathbf{n}}\!\left(\mathbf{W}^T\mathbf{k}\right)\\
  &=
  \int \!d^2\mathbf{k}\, e^{-\frac{1}{2}\mathbf{k}^T\left(\mathbf{U}\mathbf{U}^T +
    \mathbf{V}\mathbf{V}^T +
    \mathbf{W}\mathbf{W}^T\right)\mathbf{k}}\;
  Z_{\mathbf{p}}\!\left(\mathbf{U}^T\mathbf{k}\right)
  Z_{\mathbf{q}}\!\left(\mathbf{V}^T\mathbf{k}\right)
  Z_{\mathbf{n}}\!\left(\mathbf{W}^T\mathbf{k}\right)
  \label{eqn:conv-integral-split}
  \\
  \intertext{with}
  Z_{\mathbf{p}}\!\left(\mathbf{U}^T\mathbf{k}\right) &\equiv
  \Ht_{p_1}\!\left( U_{11}x_1 + U_{21}x_2 \right)\,
  \Ht_{p_2}\!\left( U_{12}x_1 + U_{22}x_2 \right) \\
  \Ht_n\!\left(x\right) &\equiv
  \left(\sqrt{\pi}\,2^n\,n!\right)^{-1/2} H_n\!\left(x\right)
  \label{eqn:normalized-hermite-def}
\end{align}
We prefer the ``normalized'' Hermite polynomials $\Ht_n$ here
over the standard Hermite polynomials $H_n$ both because they provide
a more concise notation and because their recurrence relations (see
Appendix~\ref{sec:normalized-hermite-polynomials}) are more
numerically stable \citep{NumRecipes}.

Returning to Equation \eqnref{eqn:conv-integral-split}, we can simplify the
exponential factor by requiring
\begin{equation}
  \mathbf{W}\mathbf{W}^T = \mathbf{U}\mathbf{U}^T +
  \mathbf{V}\mathbf{V}^T
  \label{eqn:conv-transform}
\end{equation}
This reduces to the familiar formula for the convolution of elliptical
Gaussians at zeroth order; for an elliptical shapelet expansion with
ellipse transform $\mathbf{S}$, $\mathbf{S}\mathbf{S}^T$ is the
covariance matrix of the Gaussian.\footnote{Note that this relation
  between the basis ellipses does not
  assume or require that the ellipticity of the functions we are
  convolving obey the Gaussian convolution rule; while the basis ellipse
  uniquely determines the ellipticity for the zeroth-order term in a
  shapelet expansion, the ellipticity of a higher-order expansion is
  also a function of the coefficients.}  We then make the
substitution $\mathbf{W}^T\mathbf{k} \rightarrow \mathbf{y}$:
\begin{align}
  I_{\mathbf{p},\mathbf{q},\mathbf{n}}\!\left(\mathbf{U},\mathbf{V},\mathbf{W}\right)
  &=
  \int \!d^2\mathbf{k}\, e^{-\mathbf{k}^T\mathbf{W}\mathbf{W}^T\mathbf{k}}\;
  Z_{\mathbf{p}}\!\left(\mathbf{U}^T\mathbf{k}\right)
  Z_{\mathbf{q}}\!\left(\mathbf{V}^T\mathbf{k}\right)
  Z_{\mathbf{n}}\!\left(\mathbf{W}^T\mathbf{k}\right) \\
  &= \frac{1}{|W|}\int \!d^2\mathbf{y}\, e^{-\mathbf{y}^T\mathbf{y}}\;
  Z_{\mathbf{p}}\!\left(\mathbf{U}^T\mathbf{W}^{-T}\mathbf{y}\right)
  Z_{\mathbf{q}}\!\left(\mathbf{V}^T\mathbf{W}^{-T}\mathbf{y}\right)
  Z_{\mathbf{n}}\!\left(\mathbf{y}\right)
  \label{eqn:conv-integral-substituted}
\end{align}

Each normalized Hermite polynomial can be represented as a linear
combination of monomials:
\begin{equation}
  \Ht_n(x) = \sum_m^N M_{n,m} x^m\;,
\end{equation}
where $M$ is the lower-triangular matrix of normalized
Hermite polynomial coefficients.  We can also write a monomial as a
linear combination of normalized Hermite polynomials using the inverse
matrix:
\begin{equation}
  x^m = \sum_n^N M^{-1}_{m,n} \Ht_n(x)\;.
\end{equation}
Along with the binomial theorem, these can be used to factor the
ellipse-transform elements out of the polynomials:
\begin{align}
  Z_{\mathbf{n}}\!\left(\mathbf{T}\mathbf{x}\right)
  &=
  \Ht_{n_1}\!\left( T_{11}x_1 + T_{12}x_2 \right)\,
  \Ht_{n_2}\!\left( T_{21}x_1 + T_{22}x_2 \right) \\
  &=
  \sum_{m1}^N \sum_{m2}^{N-m_1} M_{n_1,m_1} M_{n_2,m_2} 
  \left( T_{11}x_1 + T_{12}x_2 \right)^{m_1}\,
  \left( T_{21}x_1 + T_{22}x_2 \right)^{m_2} \\
  &=
  \sum_{m1}^N \sum_{m2}^{N-m_1} \sum_{k_1}^{m_1} \sum_{k_2}^{m_2}
  M_{n_1,m_1}M_{n_2,m_2}\binom{m_1}{k_1}\binom{m_2}{k_2}\;
  T_{11}^{m_1-k_1}\,T_{12}^{k_1}\,T_{21}^{m_2-k_2}\,T_{22}^{k_2}\,
  x_1^{m_1+m_2-k_1-k_2}\,x_2^{k_1+k_2}\\
  &=
  \sum_{m1}^N \sum_{m2}^{N-m_1} \sum_{k_1}^{m_1} \sum_{k_2}^{m_2}
  \sum_{j_1}^N \sum_{j_2}^{N-j_1} \biggl[
  M_{n_1,m_1} M_{n_2,m_2} \binom{m_1}{k_1} \binom{m_2}{k_2} 
  M^{-1}_{m_1+m_2-k_1-k_2,j_1} M^{-1}_{k_1+k_2,j_2}\biggr.\nonumber\\
  & \quad\quad\quad \biggl. \times \quad
  T_{11}^{m_1-k_1}\,T_{12}^{k_1}\,T_{21}^{m_2-k_2}\,T_{22}^{k_2}\,
   \biggr]\Ht_{j_1}\!(x_1)\,\Ht_{j_2}\!(x_2)\\
  &=
  \sum_{\mathbf{j}}^{|j|\le N}
  Q_{\mathbf{n},\mathbf{j}}\!\left(\mathbf{T}\right)
  Z_{\mathbf{j}}\!\left(\mathbf{x}\right)
  \label{eqn:hermite-transform-matrix}
\end{align}

Plugging Equation \eqnref{eqn:hermite-transform-matrix} into
Equation \eqnref{eqn:conv-integral-substituted}, we have
\begin{align}
  I_{\mathbf{p},\mathbf{q},\mathbf{n}}\!\left(\mathbf{U},\mathbf{V},\mathbf{W}\right)
  &= \frac{1}{|W|}\sum_{\mathbf{l}}^{|l|\le N_f} \sum_{\mathbf{m}}^{|m|\le N_g} 
  Q_{\mathbf{p},\mathbf{l}}\!\left(\sqrt{2}\mathbf{U}^T\mathbf{W}^{-T}\right)
  Q_{\mathbf{q},\mathbf{m}}\!\left(\sqrt{2}\mathbf{V}^T\mathbf{W}^{-T}\right)\nonumber\\
  &\quad\quad \times \quad \int \!d^2\mathbf{y}\, e^{-\mathbf{y}^T\mathbf{y}}\;
  Z_{\mathbf{l}}\!\left(\mathbf{y}/\sqrt{2}\right)
  Z_{\mathbf{m}}\!\left(\mathbf{y}/\sqrt{2}\right)
  Z_{\mathbf{n}}\!\left(\mathbf{y}\right)\\
  &= \frac{1}{|W|}\sum_{\mathbf{l}}^{|l|\le N_f} \sum_{\mathbf{m}}^{|m|\le N_g} 
  Q_{\mathbf{p},\mathbf{l}}\!\left(\sqrt{2}\mathbf{U}^T\mathbf{W}^{-T}\right)
  Q_{\mathbf{q},\mathbf{m}}\!\left(\sqrt{2}\mathbf{V}^T\mathbf{W}^{-T}\right)\nonumber\\
  &\quad\quad \times \quad \int \!d^2\mathbf{y}\, 
  \Phi_{\mathbf{l}}\!\left(\mathbf{y}/\sqrt{2}\right)
  \Phi_{\mathbf{m}}\!\left(\mathbf{y}/\sqrt{2}\right)
  \Phi_{\mathbf{n}}\!\left(\mathbf{y}\right)\\
  &= \frac{2}{|W|}\sum_{\mathbf{l}}^{|l|\le N_f} \sum_{\mathbf{m}}^{|m|\le N_g} 
  Q_{\mathbf{p},\mathbf{l}}\!\left(\sqrt{2}\mathbf{U}^T\mathbf{W}^{-T}\right)
  Q_{\mathbf{q},\mathbf{m}}\!\left(\sqrt{2}\mathbf{V}^T\mathbf{W}^{-T}\right)\nonumber\\
  &\quad\quad \times \quad 
  B_{l_1,m_1,n_1}\!\left(\sqrt{2},\sqrt{2},1\right)\;
  B_{l_2,m_2,n_2}\!\left(\sqrt{2},\sqrt{2},1\right)
  \label{eqn:convolution-complete}
\end{align}
where $B_{l,m,n}$ is the one-dimensional triple product integral defined by
\citet{RB03}; it can be computed directly with recurrence relations (see
Appendix~\ref{sec:triple-product-integral}).

The complete formula for convolution of elliptical shapelets is thus
\begin{align}
  h_{\mathbf{n}} &= 
  \frac{4\pi \left|U\right|\left|V\right|}{\left|W\right|}\!
  \sum_{\mathbf{p}}^{|p|\le N_f} \sum_{\mathbf{q}}^{|q|\le N_g}
  \sum_{\mathbf{l}}^{|l|\le N_f} \sum_{\mathbf{m}}^{|m|\le N_g} 
  i^{|p|+|q|-|n|}\;
  Q_{\mathbf{p},\mathbf{l}}\!\left(\sqrt{2}\mathbf{U}^T\mathbf{W}^{-T}\right)
  Q_{\mathbf{q},\mathbf{m}}\!\left(\sqrt{2}\mathbf{V}^T\mathbf{W}^{-T}\right)\nonumber\\
  &\quad\quad\times\quad B_{l_1,m_1,n_1}\!\left(\sqrt{2},\sqrt{2},1\right)\;
  B_{l_2,m_2,n_2}\!\left(\sqrt{2},\sqrt{2},1\right)
\end{align}

Because $B_{l,m,n}\!\left(\sqrt{2},\sqrt{2},1\right)$ is zero for
$n>l+m$, this relation is exact if $N_h \ge N_f + N_g$ and $W$ is
chosen as the solution to equation \eqnref{eqn:conv-transform},
allowing an elliptical shapelet galaxy model to be convolved with an
elliptical shapelet PSF model without approximation. 

\section{Modeling with compound shapelets}
\label{sec:modeling}

In this section we propose a simple solution to
the radial profile problem: instead of increasing the order of a
shapelet expansion in order to fit S\'{e}rsic-like profiles, we create a
compound shapelet expansion that combines multiple low-order shapelet
expansions with different radii.  By adding a low-order expansion
with small radius, a compound basis can represent a realistically cuspy
core.  Likewise, additional expansions with large radii should be able
to compactly represent extended wings without introducing
oscillatory artifacts.

In particular, we will consider compound basis functions of the form
\begin{equation}
  \Psi_n\!\left(\mathbf{x}\right|\mathbf{e}) = \sum_j^{j \le N_{\beta}} \sum_k^{k \le N_{\Phi}\![j]} a[j]_{k,n}
  \Phi_k\!\left(\left(\beta[j]\,\mathbf{S}_e\right)^{-1}\mathbf{x}\right)
  \label{eqn:compound-basis}
\end{equation}
While the compound basis thus has a single basis ellipse $\mathbf{e}$
that defines the transform $\mathbf{S_e}$, it is composed of several
shapelet bases, each with a different relative scale $\beta[j]$.  Each
compound basis function may be a linear combination of several
shapelet basis functions, with weights given by the matrices
$a[j]_{k,n}$. 

A compound basis is thus defined by a sequence of weight matrices
$a_{k,n}$ and relative scales $\beta$ (note that all the weight matrices
will have the same number of columns, but may have differing
numbers of rows).  When fitting a single galaxy, the ellipse parameters
$\mathbf{e}$ (center, complex ellipticity, and radius) will be
considered free parameters, along with the coefficients of the
compound basis functions. 

Most of the techniques we discuss below also apply to
a more general compound basis, in which the component shapelet
expansions can have entirely different ellipse parameters, but such a
basis probably has too much flexibility for most galaxy modeling
applications.  The more general expansion may be of use in PSF modeling,
however, especially if a physical model of PSF spatial variation
suggests that the ellipticity of the PSF varies differently at large
and small radii.

\subsection{Properties of the Compound Basis}
\label{sec:modeling:compound-basis-properties}

Two notable properties of the standard shapelet basis are its
orthonormality and completeness.  Both of these are in general broken
in the construction of a compound basis.

The orthonormality condition for standard shapelets
\begin{equation}
  \int_{-\infty}^{\infty}\!\phi_j(x)\,\phi_k(x)\,dx = \delta_{jk}
\end{equation}
is defined for a continuous integral with infinite limits.  While this
is an important property for deriving analytic formulae involving
shapelet basis functions, it is of limited practical use in
decomposing image data into shapelets, because the pixelization and
finite limits of image data do not match the idealized orthonormality
condition; even standard shapelet basis functions are not orthonormal
when projected to images\citep{Massey2004}.  As a result, most
standard shapelet decomposition techniques use linear least-squares
methods that do not require orthonormality of the basis
functions, and we will be able to use these same methods with little
or no modification in decomposing image data into compound shapelets.
In deriving analytic formulae for the compound basis functions, it is
already simpler to start with the analogous formulae for standard
shapelets and multiply by the weight matrices $a[j]_{k,n}$.  As we
will discuss below, however, approximate orthonormality may be useful
in some situations, and if necessary, the Gram--Schmidt procedure can
be applied to a set of compound basis functions to produce an orthonormal
compound basis.

The completeness of the standard shapelet basis also rarely useful in
practice; while an infinite shapelet expansion can perfectly represent
any function in $\mathbb{R}^2$, in practice we must work with finite
expansions.  As we have noted, a finite-size shapelet basis can be
poor at representing the particular functions--those that mimic
galaxy and PSF profiles - we are most interested in.  While the
finite-size compound basis thus lacks the
completeness properties of the infinite standard shapelet basis, our
essential argument is that a compound basis can be significantly
more complete than a similarly-sized standard shapelet basis
over the domain of functions we are most interested in.

\subsection{Determining the Basis Ellipse}
\label{sec:modeling:basis-ellipse}

For a vector of pixel values $\mathbf{z}$ and errors
$\mathbf{\sigma}$, we can solve for the best-fit ellipse parameters
and basis coefficients in a standard least-squares sense:
\begin{equation}
  \min_{\mathbf{b},\mathbf{e}} \sum_m \left[\frac{1}{\sigma_m}\left(z_m -
    \sum\limits_n\Psi_n\!\left(x_m|\mathbf{e}\right)
    b_n\right)\right]^2
  \label{eqn:least-squares}
\end{equation}
In general, however, the ellipse parameters $\mathbf{e}$ will be
highly degenerate with those linear combinations of basis coefficients
$\mathbf{b}$ that produce shapelet-space translation, shear, and
scaling transforms.  For a compound basis composed of low-order
shapelet expansions, these degeneracies are reduced, because the basis
functions poorly approximate the ellipse transform.  However, even
first- or second-order shapelet components will generally make the
model too flexible, as first-order terms approximate a
centroid shift, and second-order terms approximate changes in
ellipticity and size. 

\citet{BJ02} propose determining the ellipse parameters using a
``null test'' algorithm, in which
the parameters of the basis ellipse are iterated until the resulting
expansion has zero centroid, zero ellipticity, and unit size in the
ellipse-transformed coordinate system.  This is equivalent to solving
Equation \eqnref{eqn:least-squares} subject to the constraint
\begin{align}
  \sum_n K_{m,n} b_n &= 0
  \label{eqn:ellipse-constraint}\\
  \intertext{with}
  K_{1,n} &= \int d^2\mathbf{x}\, x_1\, \Phi_{n}\!(\mathbf{x})\; W\!\left(||x||\right)\\
  K_{2,n} &= \int d^2\mathbf{x}\, x_2\, \Phi_{n}\!(\mathbf{x})\; W\!\left(||x||\right)\\
  K_{3,n} &= \int d^2\mathbf{x}\, (x_1^2 - x_2^2)\, \Phi_{n}\!(\mathbf{x})\;
  W\!\left(||x||\right) \\
  K_{4,n} &= \int d^2\mathbf{x}\, x_1 x_2\, \Phi_{n}\!(\mathbf{x})\;
  W\!\left(||x||\right) \\
  K_{5,n} &= \int d^2\mathbf{x}\, (x_1^2+ x_2^2-1)\, \Phi_{n}\!(\mathbf{x})\;
  W\!\left(||x||\right) 
\end{align}
for some radial weight function $W$.

We can construct the analogous constraint matrix for a compound shapelet basis
simply by replacing $\Phi$ with $\Psi$ above.  Rather than applying the
constraint when fitting individual galaxies, however, we can instead
``circularize'' the basis itself by modifying the weight
matrices $\mathbf{a}$ in equation \eqnref{eqn:compound-basis}.  Given a set of
input basis functions $\Psi_n(\mathbf{x})$ 
and its constraint matrix $\mathbf{K}$, we first compute the singular
value decomposition of $\mathbf{K}^T$:
\begin{equation}
  \mathbf{U} \mathbf{D} \mathbf{V}^T = \mathbf{K}^T\;.
\end{equation}
The diagonal matrix $\mathbf{D}$ will have at most five non-zero
elements, corresponding to the first five columns of 
$\mathbf{U}$.  The remaining columns of $\mathbf{U}$ give the mapping from
the original basis to the circularized basis:
\begin{equation}
  \Psi^c_n(\mathbf{x}) = \sum_m \Psi_m(\mathbf{x}) \; U_{m,n+5} 
\end{equation}
or
\begin{equation}
  a^c[j]_{k,n} = \sum_m a[j]_{k,m} \; U_{m,n+5}\;.
\end{equation}

The resulting basis is naturally five elements smaller than the input
basis, and its constraint matrix is the zero matrix, so
Equation \eqnref{eqn:ellipse-constraint} is always true.  We can then safely
apply the least-squares Equation \eqnref{eqn:least-squares}, using the
circularized basis, to fit both the ellipse parameters and basis
coefficients simultaneously.

\begin{figure*}
    \includegraphics[width=\textwidth]{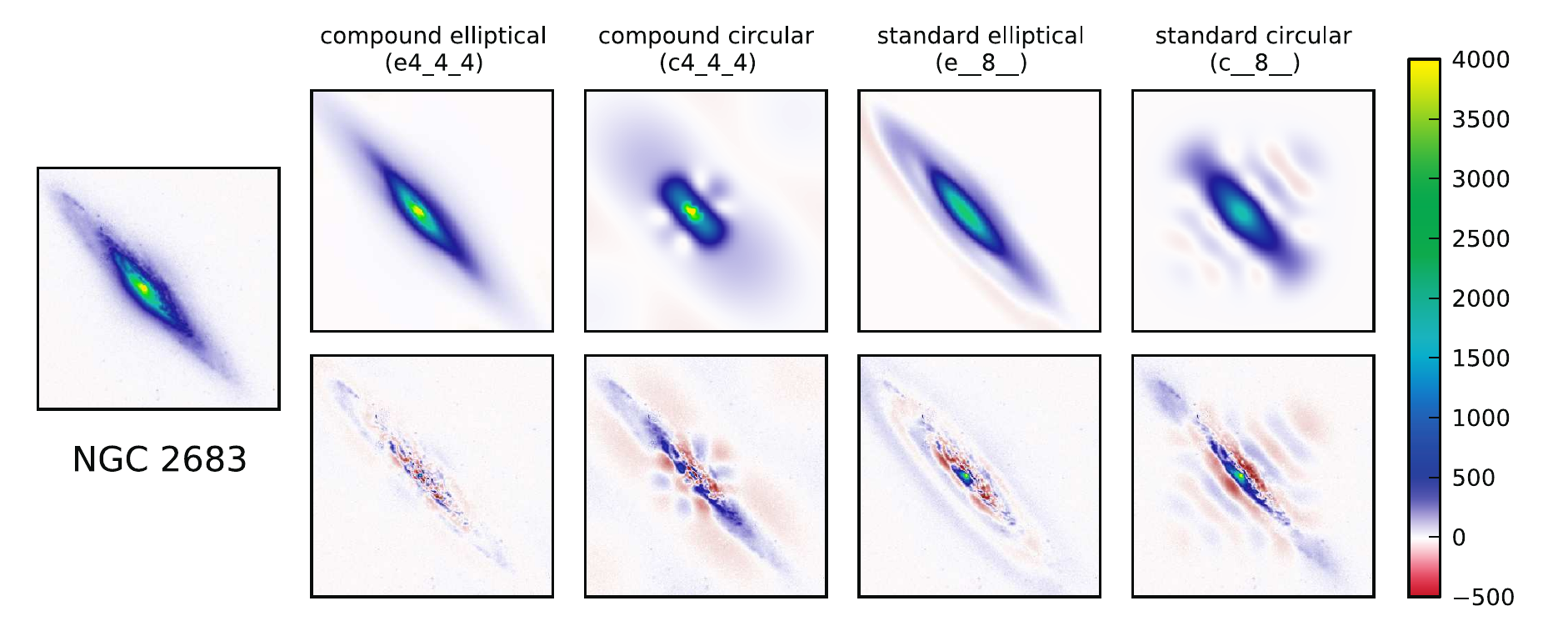}
    \caption{Image, shapelet models (upper),
      and residuals (lower) for NGC~2683, a typical edge-on spiral
      galaxy in our test sample.  Each model has 45 free parameters,
      equivalent to an 8th-order standard circular shapelet fit.
      This example clearly illustrates the artifacts
      introduced in circular shapelet fits to high-ellipticity galaxy
      images; in this case, the circular models actually have negative
      flux in certain regions.  As expected, the compound and standard shapelet fits
      perform similarly at moderate radii, but the compound
      basis is noticeably better at representing structure at small
      and large scales, modeling the core well while reducing or
      eliminating the ``ringing'' residuals present in the wings of
      the standard basis fits.  The basis labels are defined in
      Table~\ref{table:basis-sets}, and the data and fitting procedure
      are described in section~\ref{sec:demos}.
      \label{fig:ngc2683}
    }
\end{figure*}

\subsection{Radii and Shapelet Order}
\label{sec:modeling:radii-and-order}

The primary disadvantage of the compound basis technique is its
flexibility.  The user must select in advance the scales $\beta[j]$ and
shapelet orders $N_{\Phi}[j]$ to be used in fitting an ensemble of
galaxy images.  This will typically require some experimentation on a
smaller training sample of galaxies.

It is not immediately clear what metric to optimize in selecting the
scales and shapelet orders, even for an ideal training set; one does
not want to simply ensure the compound basis fits the mean morphology
well.  A better choice would be to optimize the sum of the $\chi^2$
values for all galaxies in the training sample, treating the
individual galaxy parameters as nuisance parameters.  This is a huge
global nonlinear optimization problem, however, requiring each galaxy
in the sample to be fit for a new ellipticity and set of coefficients
for every iteration in the basis parameters.  Furthermore, this
metric imposes a signal-to-noise weighting on the members of the
training sample, which may or may not be appropriate for different
training sets.  Most importantly, it is clear that a basis with a
large total number of basis functions will generally outperform a
basis with fewer, and thus part of the challenge is in determining
the correct number of basis functions to use.  There are also
computational advantages to using fewer shapelet expansions, just
as there are advantanges in decreasing the order of the shapelet
components, and a useful metric must assign weights to each of these
opposing priorities. 

One approach is to set the scales and shapelet orders to match a
particular analytic function, such as an exponential or other
fixed-index S\'{e}rsic profile.  However, to do so, one must choose a
metric that defines similarity between the analytic profile and its
shapelet approximation.  While a compound basis can approximate a
S\'{e}rsic function over a wider range of scales than a standard
shapelet basis, it still cannot reproduce a general S\'{e}rsic profile
over the full infinite range of the function, and a metric that puts
significant weight at large or very small radii typically produces a
poor approximation over the range of radii where galaxy profiles can
realistically be observed.  More importantly, even if the
analytic function to be matched is a good proxy for the mean or most
common galaxy profile, tuning the basis to fit a single fiducial
profile may reduce the flexibility of
the basis, making it worse at fitting morphologies that differ
significantly from the fiducial profile.  We have not
explored matching to analytic functions enough to rule this approach
out as an aid in determining the scales and shapelet orders, but the
difficulty in choosing a metric and the danger of producing an
inflexible basis have caused us to focus our attention on heuristic
and training-set approaches that do not involve a fiducial profile.

While we do not have a generalized method to determine optimal values for
$\beta[j]$ and $N_{\Phi}[j]$ for a sample of galaxies, a closer look
at the form of the first few shapelet basis functions provides
insight into how to construct a reasonable compound basis.  If we
consider a typical compound basis as one that contains shapelet
expansions at small, medium, and large radii, using a zeroth-order
expansion for all three will restrict the basis to models with complete
elliptical symmetry--it will be a sum of Gaussians that share the
same center and ellipticity, and differ only in size.  As we noted in
the previous section, first-order shapelet terms approximate a shift
in centroid, and as a result adding first-order terms to the component
shapelet bases of a compound basis allows elliptical
isophotes with different centers at different radii.  Likewise, adding
second-order terms allows elliptical isophotes that also vary in
ellipticity as a function of radius.

This is a crucial point in reducing the particular underfitting shear
bias noted by \citet{VB09}, as it allows the compound basis to
represent, for instance, bulge + disk galaxies in which different
components have different ellipticities.  Note that this flexibility
is not affected by the circularization procedure of the previous
section.  Circularization does not force the basis functions to have
constant ellipticity as a function of radius; it only sets (via the
weight $W$) what we define to be the mean ellipticity over all radii.

While first- and second-order shapelet components can represent some
structure beyond the shifted and sheared elliptical isophotes
discussed here, in general non-elliptical structure will be best fit
by adding higher-order terms to one or more shapelet components rather
than adding additional low-order components.  We will explore some of
these tradeoffs, and the choice of radii, in Section~\ref{sec:demos},
but we caution that these questions should generally be revisited for
different galaxy samples.

\subsection{Dimensionality Reduction}
\label{sec:modeling:dimensionality-reduction}

Because the radii and shapelet orders must be chosen in advance, a
compound basis will typically have a fixed size (particularly if it is
circularized), and unlike the standard shapelet basis, there is no
consistent way to increase or decrease the size of the basis to match
the signal-to-noise ratio (S/N) and resolution of the image data being modeled.

Given a training sample, one can attempt to mitigate some of
these problems by changing the weight matrices $\mathbf{a}$, eliminating
those linear combinations of basis functions which do not play an important
role in fitting the training sample.  This can be used to produce a
``reduced'' compound basis, in which the number of basis functions can
be much smaller than the number of shapelet basis functions used to
build it.  Quantifying the ``importance''
of a basis function is difficult, however, and may vary between
different modeling applications and datasets. 

A standard application of principal component analysis (PCA) is 
probably a poor choice.  PCA generally requires independent and
identically distributed measurement vectors, and neither is generally
true for the basis coefficient vectors $\mathbf{b}$ produced by the
fitting algorithm of Section~\ref{sec:modeling:basis-ellipse}, unless
the training sample has such high S/N and resolution that measurement
errors can be neglected.  In fact, basis coefficients will typically be
highly correlated; while the orthogonality of the standard shapelet
basis results in an approximately diagonal covariance matrix
for the fitted coefficients, compound basis functions with different
radii are not orthogonal, and will produce significant off-diagonal
terms in the covariance matrix.

In spite of this, we expect some sort of dimensionality reduction to
be a crucial part of any computationally feasible weak-lensing or
morphological analysis technique based on compound shapelets, and we
will explore this more fully in a future paper.  It may also be an
important step in using the training sample to determine the optimal
scales and shapelet orders of the component expansions; while we have
discussed the two problems separately here, constructing a basis
from a training sample involves optimizing $\beta[j]$, $N_\Phi[j]$, and
$\mathbf{a}_j$ together.

\section{Demonstration and Results}
\label{sec:demos}

To demonstrate the improvements that compound and elliptical shapelets
offer relative to more standard shapelet techniques, we apply the
compound shapelet modeling method described in the previous section to
the \citet{Frei1996} sample of nearby galaxies.  This data set has high
S/N and a very small PSF relative to the galaxy sizes, and
spans a wide range of morphologies.  Even though these conditions are
not typical of most weak-lensing observations, this is an ideal
data set for testing the suitability of our basis functions in fitting
the true, unconvolved morphologies of galaxies, even with application
to shear measurement.  While the distribution of morphologies for the
moderate- and high-redshift galaxies used in weak-lensing measurements
is not identical to the distribution of 
morphologies in our sample, the same morphological types will be
present in both, and a nearby galaxy sample affords relative image
quality that is significantly better than even the best space-based
or adaptive optics imaging of more distant galaxies.

\begin{deluxetable}{cccccc}
  \tablecaption{Shapelet order at five different radii for the nine
    demonstration basis sets.\label{table:basis-sets}} 
  \tablewidth{0pc}
  \tablehead{
    \colhead{$\beta$} & 
    \colhead{0.25} & 
    \colhead{0.50} & 
    \colhead{1.00} &
    \colhead{2.00} & 
    \colhead{4.00}
  }
  \startdata
  \texttt{\_\_8\_\_} & - & - & 8 & - & - \\
  \texttt{\_44\_4} & - & 4 & 4 & - & 4 \\
  \texttt{\_444\_} & - & 4 & 4 & 4 & - \\
  \texttt{4\_4\_4} & 4 & - & 4 & - & 4 \\
  \texttt{4\_44\_} & 4 & - & 4 & 4 & - \\
  \texttt{206\_3} & 2 & 0 & 6 & - & 3 \\
  \texttt{2063\_} & 2 & 0 & 6 & 3 & - \\
  \texttt{026\_3} & 0 & 2 & 6 & - & 3 \\
  \texttt{0263\_} & 0 & 2 & 6 & 3 & - \\
  \enddata
\end{deluxetable}

Of the 113 galaxies in the
sample, 82 were observed in $R$ and $B_j$, and the other 31 in $g$,
$r$, and $i$.  We limit our attention to the $B_j$ and $g$ images (the
two filters are very similar).
We fit each of the 113 galaxies in the sample with both the
elliptical and circular forms of each basis described below, but
discard the 31 galaxies in which the fit did not converge in one or
more cases (these failures are mostly due to large gradients in the
background present in some of the images).

\begin{figure*}
    \includegraphics[width=\textwidth]{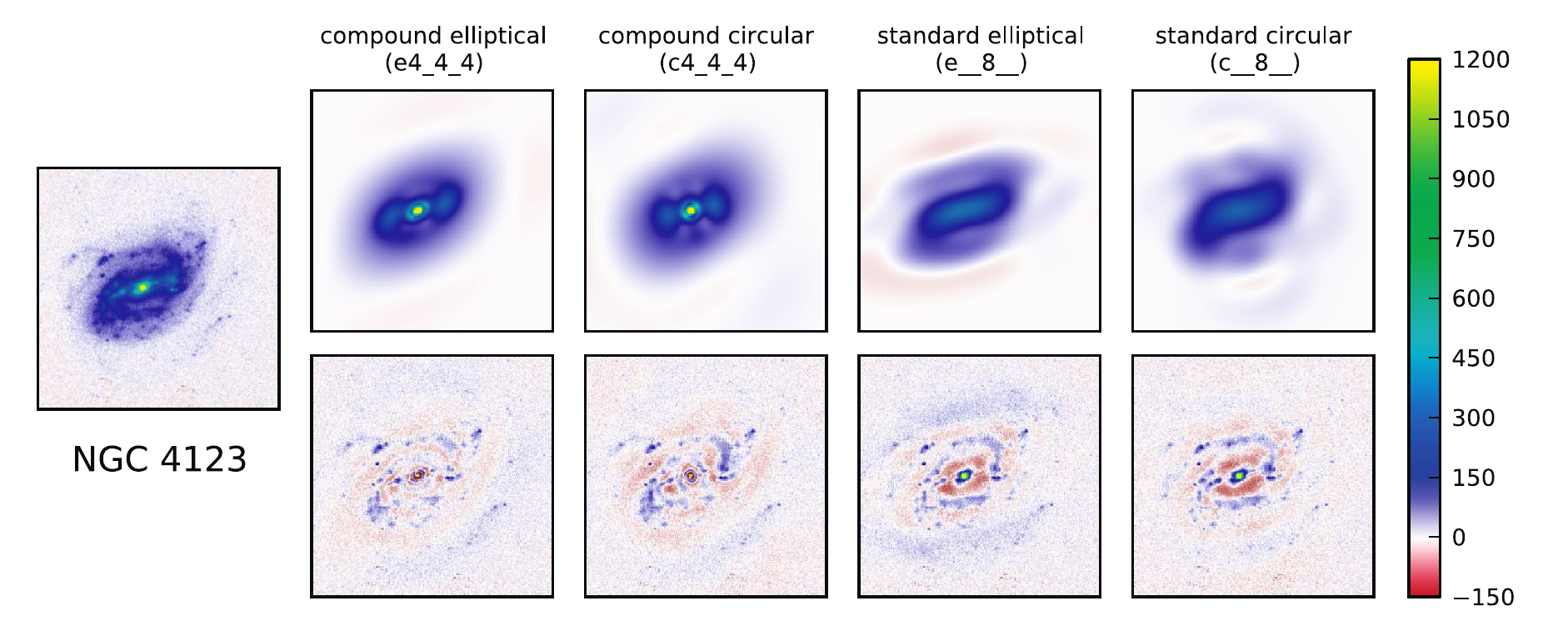}
    \caption{Image, circular shapelet models (upper),
      and residuals (lower) for NGC~4123, a typical face-on spiral
      galaxy in our test sample.  The elliptical compound basis models
      the core, bar, and ring structure reasonably well, but none of
      the models capture much of the spiral arm structure.  This is
      generally the case for the face-on spirals in our sample.
      \label{fig:ngc4123}
    }
\end{figure*}

\begin{figure*}
    \includegraphics[width=\textwidth]{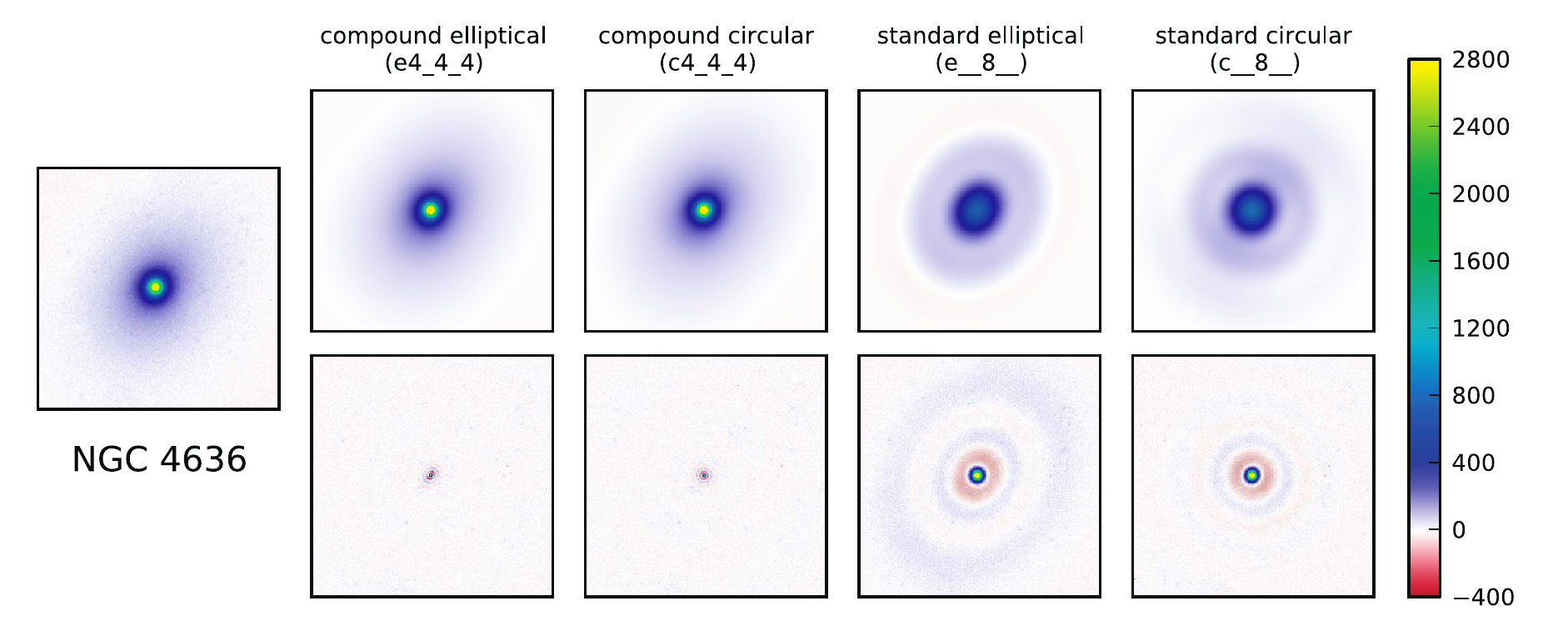}
    \caption{Image, shapelet models (upper),
      and residuals (lower) for NGC~4636, a typical massive elliptical
      galaxy in our test sample.  The difference between the compound
      basis fits and the standard basis fits is striking, particulary
      in the oscillating residuals and in the core of the galaxy,
      which is not fit at all by the standard basis.  Both circular
      fits show residuals which are clearly circular, rather than
      matched to the ellipticity of the galaxy, and the standard
      circular shapelet model is quite clearly circular at large
      radii, even though the galaxy is not; the limited basis simply
      does not have the modeling power to apply the correct shear at
      both moderate and large radii.
      \label{fig:ngc4636}
    }
\end{figure*}

We consider nine compound basis sets, including a single-scale
standard shapelet basis, each with 45 basis functions (before
circularization).  These are summarized in
Table~\ref{table:basis-sets}.  For elliptical shapelet fits
(\texttt{e} prefix), we circularize the basis with weight function
$W=1$ and fit for the coefficients and ellipse parameters 
simultaneously using the Levenberg--Marquardt nonlinear least-squares
optimizer.  The initial ellipse parameters are determined from the
unweighted moments of the image.  For circular shapelet fits
(\texttt{c} prefix), we fix
the centroid and size to the results from the elliptical fit, and
perform a linear least-squares fit using the uncircularized basis.  In
each case, there are 45 free parameters.  We use ``forward
fitting''\footnote{The PSF-convolved model is compared with the observed image at every
  iteration.}
in all cases to handle the PSF, with a simple circular Gaussian PSF
model (with FWHM as given in the image headers).

\subsection{Results}

The results for three representative galaxies are shown in
Figures~\ref{fig:ngc2683}, \ref{fig:ngc4123}, and \ref{fig:ngc4636}.
As expected, the edge-on spiral galaxy (Figure~\ref{fig:ngc2683})
demonstrates most clearly the improvement from circular to elliptical
shapelets, both for the compound basis (\texttt{4\_4\_4}) and the
standard basis (\texttt{\_\_8\_\_}).  The difference can also be seen
in the \texttt{c\_\_8\_\_} fit to the elliptical galaxy
(Figure~\ref{fig:ngc4636}).  Because the standard shapelet expansion
already has such difficulty representing the de Vaucouleur profile, it
cannot afford to ``spend'' coefficients on matching the ellipticity of
the galaxy. The compound basis provides better fits to
both the inner radii and outer radii of all the galaxies shown, and
eliminates the ``ringing'' residuals except at very small radii; this
is particularly dramatic for NGC 4636.  None of the basis sets
provides enough flexibility to fit the spiral arms of NGC 4123
(Figure~\ref{fig:ngc4123}) well, and this is generally the case for
the face-on spirals in the sample.  The compound basis fits do provide
a reasonably good approximation to the bar and ring structure, and all
the fits do a moderately good job of fitting the overall profile of
the spiral galaxies.
\begin{figure*}
  \begin{center}
    \includegraphics[width=\textwidth]{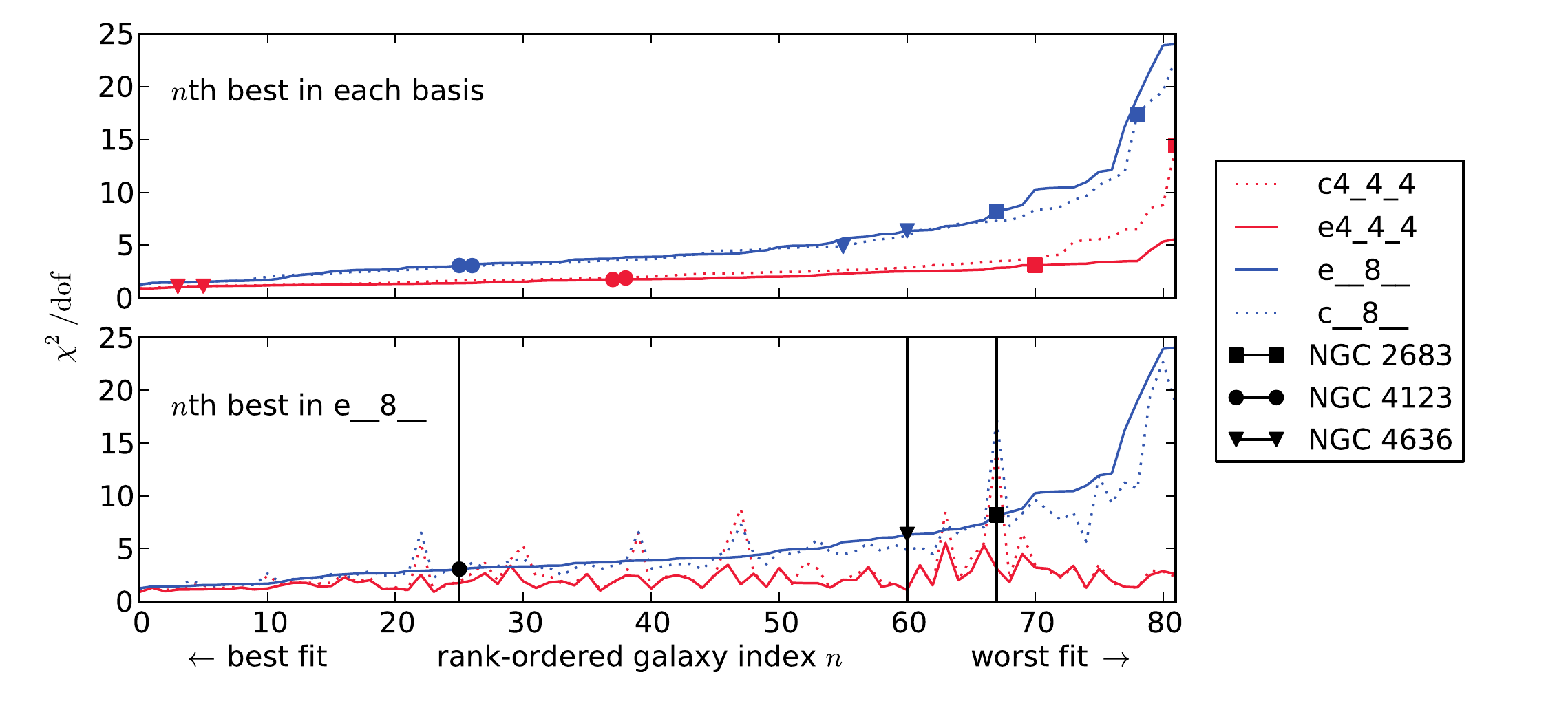}
  \end{center}
  \caption{Rank-ordered goodness of fit for 82 galaxies from the
    \citet{Frei1996} sample of nearby galaxies, for selected
    elliptical and circular shapelet basis sets.  In the upper plot,
    the galaxies are rank-ordered by reduced $\chi^2$ for each basis
    separately; each point on the $x$-axis thus corresponds to the $n$th
    best-fit galaxy for each basis.  In the lower
    plot, the reduced $\chi^2$ from the \texttt{e\_\_8\_\_} fit is
    used to set the galaxy rank index for all bases; a point on the
    $x$-axis thus corresponds uniquely to a single galaxy.  The
    goodness of fit
    for the galaxies shown in Figures~\ref{fig:ngc2683}, \ref{fig:ngc4123},
    and \ref{fig:ngc4636} are given by the square, circle, and triangle
    points, respectively.  The circular standard basis actually
    produces slightly lower average and worst-case residuals than the
    elliptical standard basis, but the compound elliptical basis is
    significantly better than either.  The high-ellipticity edge-on spirals
    (such as NGC 2683) that present a particular problem for the
    circular shapelet expansion can be clearly seen in the lower plot
    as spikes in the circular basis
    goodness-of-fit. \label{fig:residuals-ec}}  
\end{figure*}

These tendencies are quantified in Figure~\ref{fig:residuals-ec},
which shows the goodness-of-fit distribution for the 82
galaxies successfully fit with all basis sets.  The elliptical compound
basis consistently has the 
lowest residuals, and even the circular compound basis generally
outperforms both the elliptical and circular standard shapelet basis.
The elliptical standard shapelet basis is not uniformly better
than the circular standard shapelet basis, but the cases where it is --
high-ellipticity edge-on spirals, like NGC 2683 -- stand out clearly as
spikes in the lower plot, in which the galaxy index is consistent
across all the fits.  It is also worth noting that when the compound
basis is used, featureless, low-ellipticity early-type galaxies like
NGC 4636 have some of the lowest residuals in the sample.  In
contrast, these have moderately large residuals compared to the rest
of the sample for the standard shapelet basis fits, demonstrating that
using a compound basis essentially solves the problem of fitting de
Vaucouleur profiles with shapelets.

\begin{figure*}
  \begin{center}
    \includegraphics[width=\textwidth]{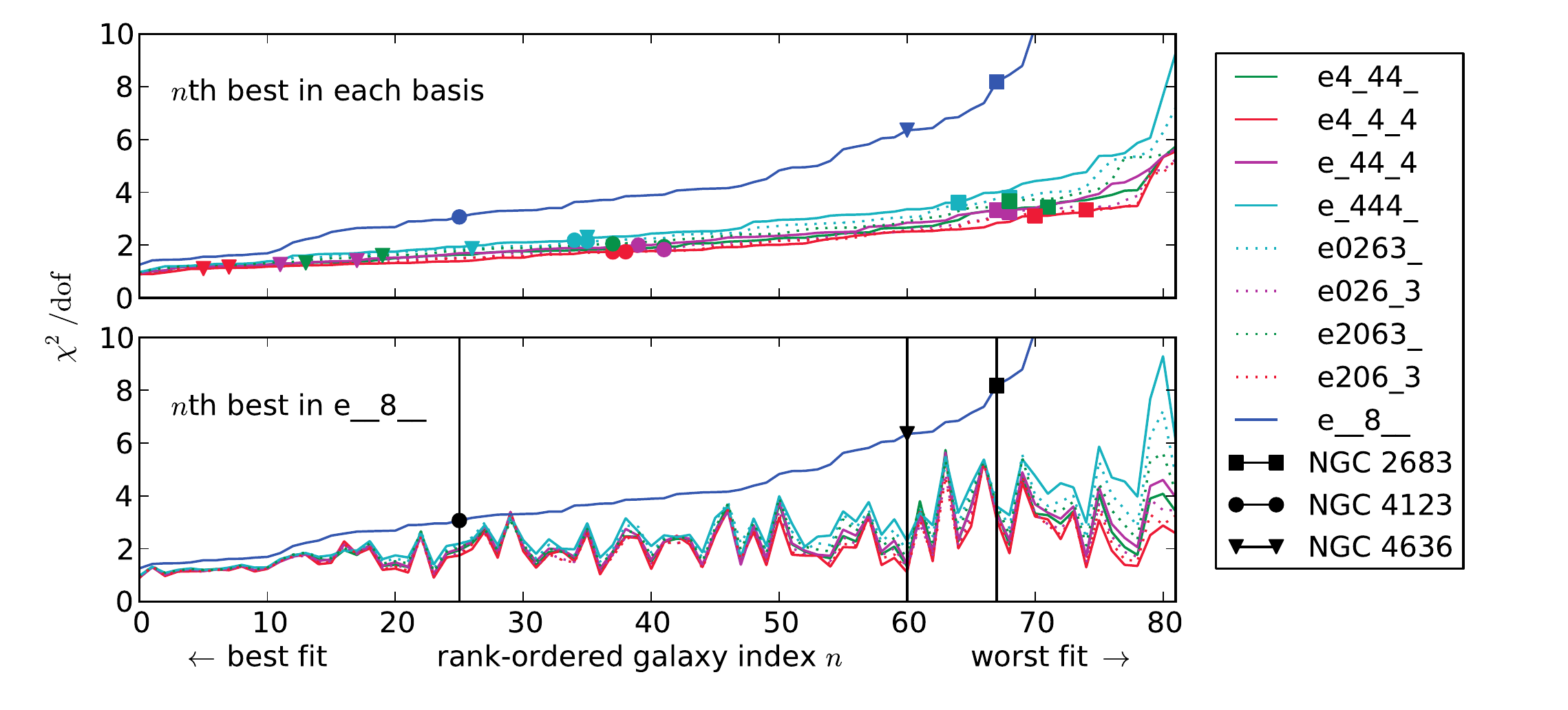}
  \end{center}
  \caption{Goodness-of-fit comparison for all elliptical
    basis sets. The rank-ordered indexing and galaxy sample are the
    same as in Figure~\ref{fig:residuals-ec}, but the $y$-axis scale
    has been changed to show more detail.  All of the compound
    shapelet basis sets outperform the standard shapelet basis for all
    galaxies but one, in which the results are comparable, and the
    compound bases show considerably better worst-case performance.
    Between compound basis sets, those with components at both the
    largest and smallest radii (\texttt{4\_4\_4} and \texttt{206\_3})
    consistently have the lowest residuals. \label{fig:residuals}}
\end{figure*}

Figure~\ref{fig:residuals} similarly shows the distribution
goodness-of-fit for all elliptical basis sets.  The most obvious
result is that all the 
compound shapelet bases tested outperform the standard shapelet basis,
both in aggregate and (with one slight exception) for every galaxy
individually.  Between the compound 
basis sets, those with shapelet expansions at both $\beta=0.25$ and
$\beta=4.0$ tend to perform slightly better, especially compared to
basis sets that contain neither, suggesting that compound bases are
most effective when constructed with shapelet orders at radii that
differ by more than a factor of two.  This provides a partial answer
to the question of how to choose the radii and orders of the component
shapelet bases that is both frustrating and somewhat comforting: while
there is no clear choice for an optimal combination of radii and
orders, any reasonable choice that samples the radial range of
significant morphology is still a marked improvement over the
standard shapelet basis.

\section{Conclusions and Future Work}
\label{sec:conclusion}

Despite the problems discussed in Section~\ref{sec:limitations},
shapelets have become one of the most important tools in galaxy
modeling, particularly in weak lensing.  The inability of the standard
shapelet basis to 
represent morphologies with high ellipticity or steep profiles puts
limits on the accuracy of shapelet-based shear estimators, however,
and likely degrades the performance of shapelet-based morphological
analyses.  We have presented here two techniques to address these
limitations. 

The first is a
convolution relation for elliptical shapelets.  This eliminates the
need to use the lossy shapelet-space shear operator, and allows
ellipse-parameterized shapelets to be used in place of a circular basis
parameterized only by centroid and radius.  While the elliptical basis
requires five nonlinear parameters to be fit instead of three,
the two additional parameters more than pull their weight in modeling
power, particularly for high-ellipticity galaxies.  Perhaps most
importantly, an exact elliptical shapelet convolution allows
shapelet-based shear estimators to be constructed that do not suffer
from the ``shear artifact'' bias discussed in
Section~\ref{sec:limitations:ellipticity}.  
The results of Section~\ref{sec:demos} do not demonstrate the advantages
of the elliptical convolution formula;
while we used it in fitting the galaxies, the size of our test
galaxies relative to the PSF makes the difference between
our exact convolution and the approximation of \citetalias{NB07}
negligible.  However, this formula 
can be immediately put to use in existing elliptical shapelet
shear--measurement methods, such as that of \citetalias{NB07},
eliminating one source of bias for galaxies near the resolution
limit.

To address the difficulties of the standard shapelet basis in fitting
galaxies with high S\'{e}rsic indices, we have introduced the concept of a
``compound'' shapelet basis -- a basis composed of multiple low-order
shapelet expansions with different scale radii.  Even with simple, ad hoc
choices for the radii and orders of the component shapelet expansions,
the compound basis is significantly better than a single higher-order
shapelet expansion at fitting realistic galaxy morphologies,
particularly in its worst-case performance.  By combining
low-order shapelet basis functions at multiple radii, a compound basis
can compactly represent both the sharp cores and extended wings of galaxies
with high S\'{e}rsic indices.  As we have seen, this can make a
significant difference even in spiral galaxies with relatively low
S\'{e}rsic indices, and almost completely eliminates the oscillatory
artifacts often present in standard shapelet fits. 
This should reduce the underfitting shear bias
for shapelet-modeling shear measurement methods, but whether the
bias can be reduced to acceptable levels for future surveys is a
question we reserve for a future paper.  Morphological analyses such as
those of \citet{KellyMcKay2004} and \cite{Andrae2010} should also benefit from
utilizing compound shapelet basis sets, as the more compact compound
shapelet representation essentially gives classification and
clustering algorithms a head start in reducing the dimensionality of
the problem.  

Unlike the elliptical convolution formula, further work is needed to
make full use of the compound shapelet concept in a complete shear
measurement or morphological analysis method.  While we have
demonstrated success in 
fitting nearby galaxies with unoptimized basis sets and a simple
fitting algorithm, more challenging modeling problems may require
techniques that tune the parameters and weights of a compound basis
using a training sample of galaxies.  Such a basis of
``eigenmorphologies'', built from analytically tractable shapelet
building blocks, would be an extremely powerful tool, not only for
morphological analysis, but also in further quantifying and reducing
underfitting biases in model-based shear measurement techniques.

\acknowledgements

The author thanks Tony Tyson for very helpful comments on
earlier drafts of this paper, without which several figures would have
been nearly incomprehensible.

This material is based upon work supported under a National Science
Foundation Graduate Research Fellowship.

\bibliographystyle{custom2}
\bibliography{citations}

\begin{appendix}
  \section{Normalized Hermite Polynomials}
  \label{sec:normalized-hermite-polynomials}
  The normalized Hermite polynomials are simply the product of the
  standard Hermite polynomials with the Gauss--Hermite normalization factor
  \begin{equation}
    \Ht_n\!\left(x\right) \equiv
    \left(\sqrt{\pi}\,2^n\,n!\right)^{-1/2} H_n\!\left(x\right).
    \tag{\ref{eqn:normalized-hermite-def}}
  \end{equation}

  Like the standard Hermite polynomials, these can be efficiently
  evaluated using recurrence relations
  \begin{equation}
    \Ht_0(x) = \pi^{-1/4}
  \end{equation}
  \begin{eqnarray}
    \Ht_{n}(x) &=& x\sqrt{\frac{2}{n}}\Ht_{n-1}(x)
    -\sqrt{\frac{n-1}{n}}\Ht_{n-2}(x) \\
    &=& x\sqrt{\frac{2}{n}}\Ht_{n-1}(x)-\frac{1}{\sqrt{2n}}\D{\Ht_{n-1}(x)}{x}
  \end{eqnarray}
  \begin{equation}
    \D{\Ht_{n}(x)}{x} = \sqrt{2n}H_{n-1}(x)
  \end{equation}
  These recurrence relations are more numerically stable than those
  for the standard Hermite polynomials, which suffer from round-off
  error at high order.

  \section{Triple Product Integral}
  \label{sec:triple-product-integral}
  
  In Section \ref{sec:convolution}, we defined the elliptical shapelet
  convolution formula \eqnref{eqn:convolution-complete} in terms of
  the \emph{triple product integral} 
  \begin{equation}
    B_{l,m,n}(\alpha,\beta,\gamma)
    = \frac{1}{\sqrt{\alpha\beta\gamma}}\int^{\infty}_{-\infty}\!\phi_l(x/\alpha)\,
    \phi_m(x/\beta)\,\phi_n(x/\gamma)\,dx
  \end{equation}
  with $\alpha=\beta=\sqrt{2}$ and $\gamma=1$.  The recurrence
  relation given by \citet{RB03} to compute this integral suffers from
  the same round-off error problems as the standard Hermite
  polynomial recurrence relations.  Using the normalized Hermite
  polynomials, we present here a more stable relation.  Let
  \begin{equation}
    B_{l,m,n}(\alpha,\beta,\gamma)
    = \frac{\sqrt{2}v}{\sqrt{\alpha\beta\gamma}}\Lt_{l,m,n}(a,b,c),
  \end{equation}
  with
  \begin{equation}
    v^{-2}=\alpha^{-2}+\beta^{-2}+\gamma^{-2},\;
    a=\sqrt{2}\frac{v}{\alpha},\:
    b=\sqrt{2}\frac{v}{\beta},\:
    c=\sqrt{2}\frac{v}{\gamma}
  \end{equation}

  \begin{equation}
    \Lt_{0,0,0} = \pi^{-1/4}
  \end{equation}
  \begin{eqnarray}
    \Lt_{l,m,n}
    &=& \sqrt{\frac{l-1}{l}}(a^2-1)\Lt_{l-2,m,n}
    + \sqrt{\frac{m}{l}}\,ab\,\Lt_{l-1,m-1,n} 
    + \sqrt{\frac{n}{l}}\,ab\,\Lt_{l-1,m,n-1} \\
    &=& \sqrt{\frac{l}{m}}\,ab\,\Lt_{l-1,m-1,n} 
    + \sqrt{\frac{m-1}{m}}(b^2-1)\Lt_{l,m-2,n}
    + \sqrt{\frac{n}{m}}\,bc\,\Lt_{l,m-1,n-1} \\
    &=& \sqrt{\frac{l}{n}}\,ac\,\Lt_{l-1,m,n-1}
    + \sqrt{\frac{m}{n}}\,bc\,\Lt_{l,m-1,n-1} 
    + \sqrt{\frac{n-1}{n}}(c^2-1)\Lt_{l,m,n-2}
  \end{eqnarray}
  
  Note that while our $B$ is the same as that defined by
  \cite{RB03}, our $\Lt$ is not the same as their $L$:
  \begin{equation}
    \Lt_{l,m,n}(a,b,c) = \left(2^{l+m+n}\,l!\,m!\,n!\,\pi^{3/2}\right)^{-1/2}\,
    L_{l,m,n}(a,b,c)
  \end{equation}

\end{appendix}

\newpage

\end{document}